\documentstyle[preprint,aps]{revtex}
\begin{document}
\draft
\title{Quantum signatures of chaos in the dynamics of a  trapped ion}
\author{J.K. Breslin~$^\dagger$, C. A. Holmes~$^\ddagger$ and G.J. 
Milburn~$^\dagger$}
\address{\dag Department of Physics,\\ \ddag Department of Mathematics\\
 The University of Queensland, QLD 4072,
Australia.\\ TEL 61-7-33653401}
\date{\today}
\maketitle
\begin{abstract}
 We show how a nonlinear chaotic system, the parametrically kicked 
nonlinear oscillator, may be realised in the dynamics of a trapped,
laser-cooled
ion, interacting with a sequence of standing wave pulses. 
Unlike the original optical scheme [G.J.Milburn and C.A.Holmes, 
Phys. Rev A, {\bf 44}, 4704, (1991) ],
the trapped ion enables  strongly quantum dynamics  with minimal
dissipation. 
This should permit an experimental test of one of the quantum signatures
of
chaos; 
irregular collapse and revival dynamics of the average vibrational
energy.
\end{abstract}

It is now well understood that the quantum dynamics of a classically
chaotic
system
will show major departures from the classical motion on a suitable 
time scale\cite{Chaos1,Chaos2}. Systems which classically show chaotic
diffusion
of a slow 
momentum-like variable, will cease to show diffusion after the
'break-time'. 
In this region, the momentum distribution is found to be exponentially
localised,
rather than Gaussian. One system which demonstrates this phenomenon
is a laser cooled atom moving under the dipole force of a 
standing wave with periodically modulated nodal positions
\cite{Graham92,Dyrting96}
Recently this prediction achieved
experimental confirmation in the work of Raizen and
coworkers\cite{Raizen94}. 
The quantum and classical dynamics can also differ through the 
existence of quantum tunnelling between fixed points of the classical 
Poincar\'{e} section, a phenomenon which has not yet been observed. In
this
paper 
we consider yet another way in which quantum and classical motion can
differ for
systems
which classically do not show diffusion. In this case the transition to 
chaos in
the classical system is manifested in the quantum case by a 
transition from regular collapse and revivals of average values with
time, to a highly irregular collapse and revival sequence\cite{Haake}. 
An example of such a system is the parametrically kicked 
nonlinear oscillator\cite{MilHolmes91}. A nonlinear oscillator, in
which the frequency is linearly dependent on energy, is periodically
kicked
by a parametric amplification process, which momentarily turns 
the origin into a hyperbolic fixed point. The classical dynamics of this
system 
exhibits regions of regular and chaotic motion in the Poincar\'{e}
section. 

In reference \cite{MilHolmes91} it was proposed to realise this system
using a
combination of a Kerr optical nonlinearity and an optical parametric
amplifier.
Unfortunately the required optical nonlinearity is usually rather small
and 
accompanied by a large amount of dissipation. For this reason, it is
unlikely
that
an optical version will ever show interesting quantum features. However
the
system
does show some interesting classical chaotic behaviour including a
strange 
attractor in which unstable periodic orbits can be stabilised by OGY
control\cite{Holmes94}.
In this paper we show that a much more promising realisation of the 
quantum version of this system may be achieved in the dynamics of a
single
trapped 
laser cooled ion interacting with a sequence of pulsed optical fields.
Furthermore
the dynamics may be monitored with high quantum efficiency using the QND
vibrational energy measurement scheme proposed by Filho and
Vogel\cite{Filho96}
together with the now standard method of detection on a strong probe
transition.

We consider the vibrational motion
of a single trapped ion. In the absence of laser pulses, the ion
executes
simple harmonic motion in the quadratic trap potential. As we show
below, 
a suitable sequence of laser pulses enables us to realise the nonlinear 
quantum dynamics specified by the map
\begin{equation}
|\psi_{n+1}\rangle = U_{NL}U_{PA}|\psi_n\rangle \label{map}
\end{equation}
where
\begin{equation}
U_{NL} = \exp\left (-i\theta a^\dagger a-i\frac{\mu}{2}(a^\dagger)^2
a^2\right )
\end{equation}
describes an intensity dependent rotation in the oscillator phase plane
while
\begin{equation}
U_{PA} = \exp\left (\frac{r}{2}((a^\dagger)^2-a^2)\right )
\end{equation}
describes a parametric amplification kick. In these expressions the
operators 
$a, a^\dagger$ are the ladder operators for the quantised harmonic
motion 
of the trapped ion, which may be written in terms of the dimensionless 
position and momentum operators $\hat{X}_1, \hat{X}_2$, by
$a=\hat{X}_1+i\hat{X}_2$.
\section{Classical dynamics}
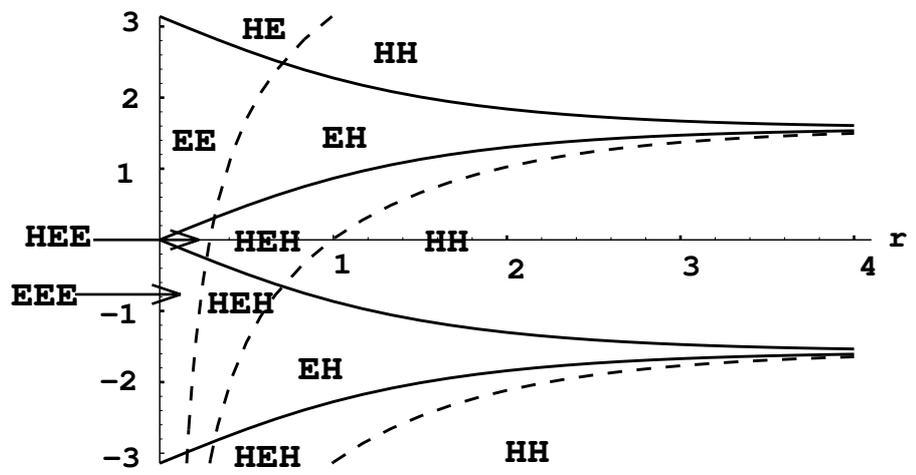
\begin{figure}
\caption{Bifurcation diagram for the origin and the period-1 orbits close to the
origin. The bifurcation curves for the origin are represented as solid
lines while those for the period-1 orbits are dashed. The regions are
labeled by a finite sequence of $E$ and $H$s, standing for elliptic and
hyperbolic. The first letter refers to the stability of the origin, the
second to that of the pair of period-1 orbits closest to the origin in
the 2nd and 4th quadrants (those in the 1st and 3rd are always
hyperbolic), the third to the pair of period-1 orbits next closest to
the origin etc.}
\label{fig1}
\end{figure}

The  classical model corresponding to Eq(\ref{map}) is closely related
to 
the model in reference \cite{MilHolmes91},
\begin{eqnarray}
X_1^\prime = g\left (\cos(\theta +\mu R^2)X_1+\sin(\theta + \mu
R^2)X_2\right
)\\
X_2^\prime = \frac{1}{g}\left (\cos(\theta + \mu R^2)X_2-\sin(\theta
+\mu
R^2)X_1\right )
\end{eqnarray}
where $R^2=X_1^2+X_2^2$ and the parametric gain is $g=e^r$. The 
parameter $\mu$ is simply a scaling parameter and the crucial control
parameters are the parametric gain $g$ and $\theta $. When $\theta=2\pi$, easily
achievable in experiment as we show below,  this model 
reduces to that discussed in reference \cite{MilHolmes91}. The origin is
then a saddle. If however
 $|\cosh r\cos\theta|\ <\ 1$, where $r=ln
g$, the origin is elliptically stable.
 The 
bifurcation diagram is shown in figure \ref{fig1}. The 
origin is elliptically stable in the regions bounded by the solid curves 
$\theta=\pm\arccos\left (\frac{1}{\cosh r}\right ) +n\pi$ and labelled
$E**$. Away from the origin 
an infinite number of period-1 orbits lie on lines with a slope of
$\pm e^{(-r)}$, 
independent of $\theta$ as in the 
original model. However the radius at which the period-1 orbits lie does 
depend on $\theta$, and is determined by the equation
\begin{equation}
-\tan(\theta+\mu R^2)=\pm \sinh r
\end{equation}
 As $\theta$ is increased the
period-1 orbits move inwards collapsing on the origin and changing its
stability. Those in the odd quadrants are always unstable. In the even
quadrants a finite number (maybe zero) lying closest to the origin are stable
depending on whether
\begin{equation}
 1 \ >\mu R^2 \tan(\theta+\mu R^2) \ >0
 \end{equation}
  They loose stability on the curves $\theta=\arctan(\sinh r)
 -\frac{1}{\sinh r} +n\pi$, shown as dashed (for n=-1,0,1) in figure
 \ref{fig1}. The various parameter regions are labelled with a sequence
 of $E$s and $H$s standing for elliptic or hyperbolic. The first letter
 refers to the stability of the origin and the subsequent letters to the
 stability of the pairs of period-1 orbits in the 2nd and 4th quadrants starting
 with those closest to the origin and moving out. Since the elliptic
 period-1 orbits always occur closest to the origin in the region
 $EEH$,for instance, 
 only one pair of period-1 orbits are elliptically stable.

In figure \ref{fig2}
we show typical Poincar\'{e} sections of the classical dynamics for
various values of 
$g$, and $\theta =2\pi$.  Note the transition to almost global chaos for
large values of $g$. 

\section{Quantum dynamics}
To compare the 
quantum and classical dynamics we compute the mean energy as a function
of the 
number of kicks. In the quantum case we assume the oscillator is
initially 
prepared in the vibrational ground state, while in the classical case we
use an 
initial Gaussian phase-space distribution with the same moments as the 
initial quantum state. The detailed results are given in reference 
\cite{MilHolmes91}, 
which we summarise here.  In the classical case, the average energy
saturates at
an almost time independent value, as the initial distribution becomes
spread
around
the period-one fixed points. The time taken to reach the saturation
value is
longer
for an initial state in the regular region, than for one located in a
chaotic
region. 
In the quantum case, the mean energy follows the classical for a short
time,
but instead of saturating at an almost stationary value, continues to
oscillate
with a characteristic collapse and revival envelope. The transition from
regular
to
chaotic dynamics in the classical system is marked in the quantum system
by a
transition from a regular collapse and revival sequence to a highly
irregular
and 
more rapid collapse and revival sequence. This is due to the fact that a 
state localised in the chaotic region has support on very many
quasi-energy
eigenstates, compared to a state localised in a regular region.

We now show how this system may be realised in a trapped ion experiment. 
Following Filho and Vogel\cite{Filho96} we first consider an ion trapped
at an
antinode 
of an optical standing wave tuned to the atomic frequency, the carrier 
frequency.  In an interaction picture
at frequency $\nu$ the interaction Hamiltonian is, 
\begin{equation}
H_I=-\hbar\Omega\eta^2 a^\dagger a\sigma_x +
\hbar\frac{\Omega\eta^4}{4}(a^\dagger)^2a^2\sigma_x
\label{ham}
\end{equation}
where $\Omega$ is the Rabi frequency and $\eta$ is the Lamb-Dicke 
parameter. 
If the ion is first prepared in an eigenstate of $\sigma_x$ by the
application
of a 
$\pi/2$ pulse, it will remain in this state. In that case the
vibrational
motion experiences a linear and a nonlinear phase shift.  The
interaction 
is precisely 
the kind required to realise $U_{NL}$, with $\theta= T\Omega\eta^2$
$\mu =\frac{T\Omega\eta^4}{2}$
where $T$ is the time for which the standing wave pulse is applied. 

In order to realise the parametric kick we use a Raman interaction with
two laser pulses, one tuned to the first lower sideband, the other to
the first 
upper sideband. This scheme has been successfully used by Wineland and
coworkers
to 
prepare the ion in a squeezed vibrational state\cite{Meekhof96}. In
this case
the interaction Hamiltonian, in an interaction picture at frequency
$\nu$,  is 
\begin{equation}
H_{PA} = i\hbar \kappa (a^2-(a^\dagger)^2)
\end{equation}
where $\kappa=\frac{\Omega_1\Omega_2\eta_R^2}{8\delta_1\delta_2}$
and $\Omega_i$ is the Rabi frequency for each of the Raman pulses, each of 
which is detuned from the atomic transition by $\delta_i$ such that  
$\delta_1-\delta_2=2\nu$. The Lamb-Dicke parameter for the Raman 
transition is $\eta_R=\delta kx_0$, where $\delta k$ is the wave vector 
difference for the two Raman beams and $x_0$ is the rms fluctuations of 
position in the ground state of the trap.  
Note that this Hamiltonian is independent of the ion electronic state. 
The ion evolves according to the parametric kick transformation with
$
g=e^{\kappa T}
$ for a Raman pulse of duration $T$.  In
summary, we first 
prepare the ion electronic state by the application of a $\pi/2$ pulse,
then apply the pulse at the carrier frequency, followed by
the Raman parametric kick pulses. Pulses at the carrier frequency and 
the Raman pulses then alternate to effect each kick.

A particular advantage of this realisation is that it is easy to read
out 
the energy of the vibrational motion at any stage. If we 
reprepare the ion in the ground electronic state at the end of the
carrier 
frequency pulse, its evolution will be dominated by the first term in 
Eq(\ref{ham}). This is a QND measurement interaction for the vibrational 
quantum number. It causes the Bloch vector describing the two level 
system to precess around the x-axis by an angle proportional to 
$a^\dagger a$. The probability to find the atom in the ground state a 
time $\tau$ after the last squeezing pulse is
\begin{equation}    
P_g(\tau)=\sum_{n=0}^\infty P(n)\cos^2 \phi\tau n
\end{equation}
where $\phi=\Omega\eta^2/2$. A set of readouts of the atomic state a
time
after
an appropriate number of kicks thus enables a moment of the photon
number to be
constructed
directly. This is similar to the scheme used by Wineland and 
coworkers to observe the Jaynes-Cummings collapse and revival
sequence\cite{Meekhof96}. 

In figures \ref{fig3} and \ref{fig4} we plot $P_g(\tau) = 
\langle \cos^2\phi\tau \hat n\rangle$ for 
the same two initial states and parameter values as discussed in
reference
\cite{MilHolmes91}. 
In figure \ref{fig3}, the initial state is localised in a classically
regular region of
phase space.
A regular collapse and revival sequence is evident in the quantum case 
(fig. \ref{fig3}a). In figure \ref{fig4} however,
we have the case of an initial state localised in the 
chaotic part of the classical phase space, leading to an irregular
revival sequence. In both cases the corresponding classical moment is shown as
well. 

We can estimate reasonable values of the parameters $\theta,g,\mu$ from
recent experiments on trapped ions. From Meekhof et al.\cite{Meekhof96}, 
typical values are $\Omega=10^6,\ \ \eta=0.2,\ \ g=6.$ For an interaction 
time of $T=10\mu\mbox{s}$, we have $\theta\approx 1.0,\ \mu\approx0.02$. 
It is then apparent that the parameter values of interest are achievable 
in an experiment.

\begin{figure}
\caption{Phase space portraits for the classical map. In all cases 
$\mu = 0.01 \pi$, $\theta = 2 \pi$. (a) $g = 1.0$, (b)$g = 1.2$, 
(c) $g = 1.5$, and (d)$g = 2.0$.}
\label{fig2}
\end{figure}

\begin{figure}
\caption{Initial states centered in a classically regular region: 
(a) Plot of the quantum average $P_g(\tau) = \langle 
\cos^2\phi\tau \hat n\rangle$ for an initial coherent state. 
(b) Plot of $\cos^2(\phi\tau E)$ (where E is the classical mean energy) 
for an initial
Gaussian distribution. In both cases the initial state is centered at
(0,0),
with $g = 1.2$, 
$\mu = 0.01 \pi$, $\theta = 2 \pi$, and $\phi\tau = 0.01$.}
\label{fig3}
\end{figure}

\begin{figure}
\caption{Same as Fig. (\ref{fig3}) except initial states are centered in
a 
classically chaotic region, at (1,0),
with $g = 1.5$.}
\label{fig4}
\end{figure}
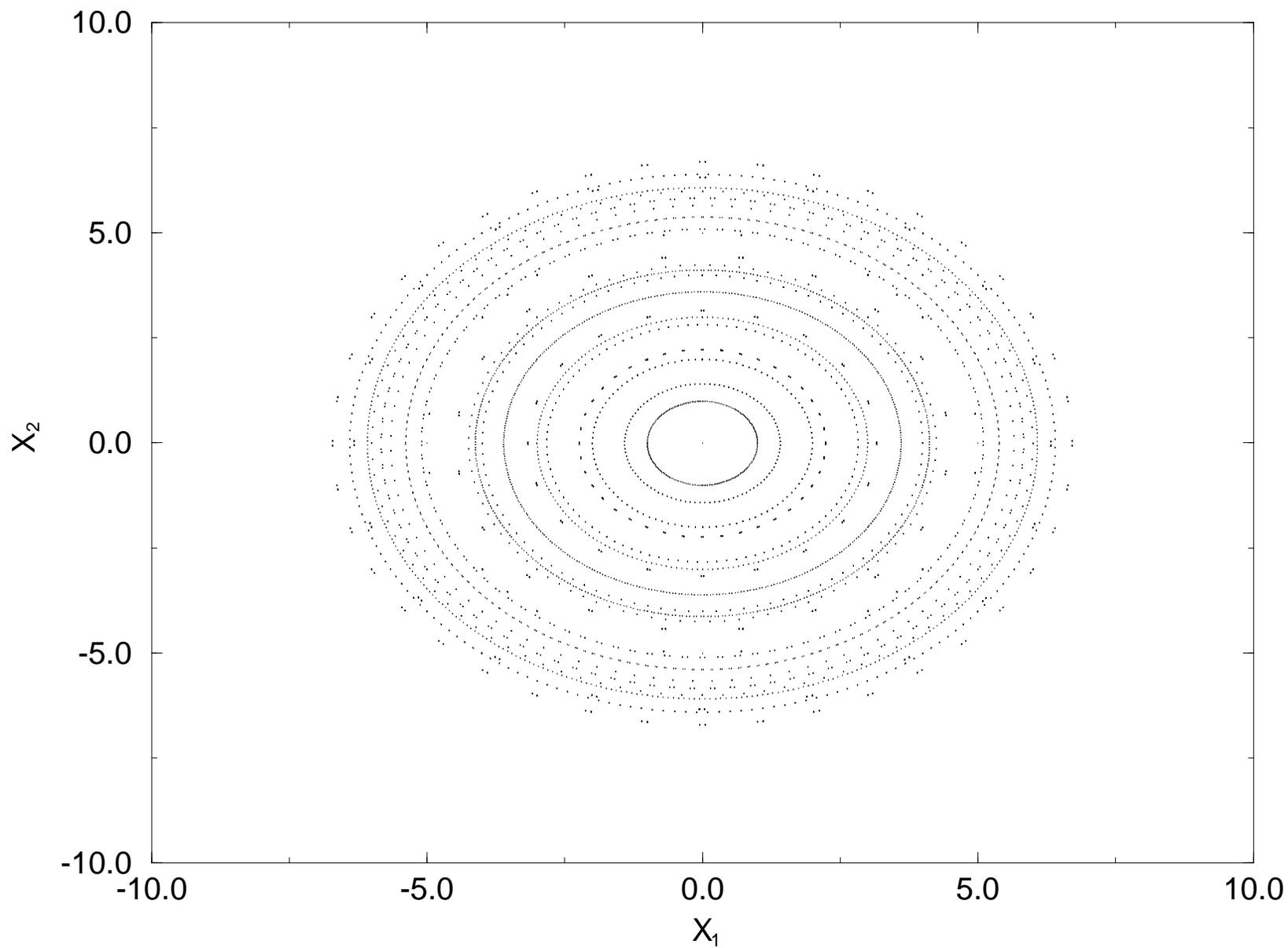
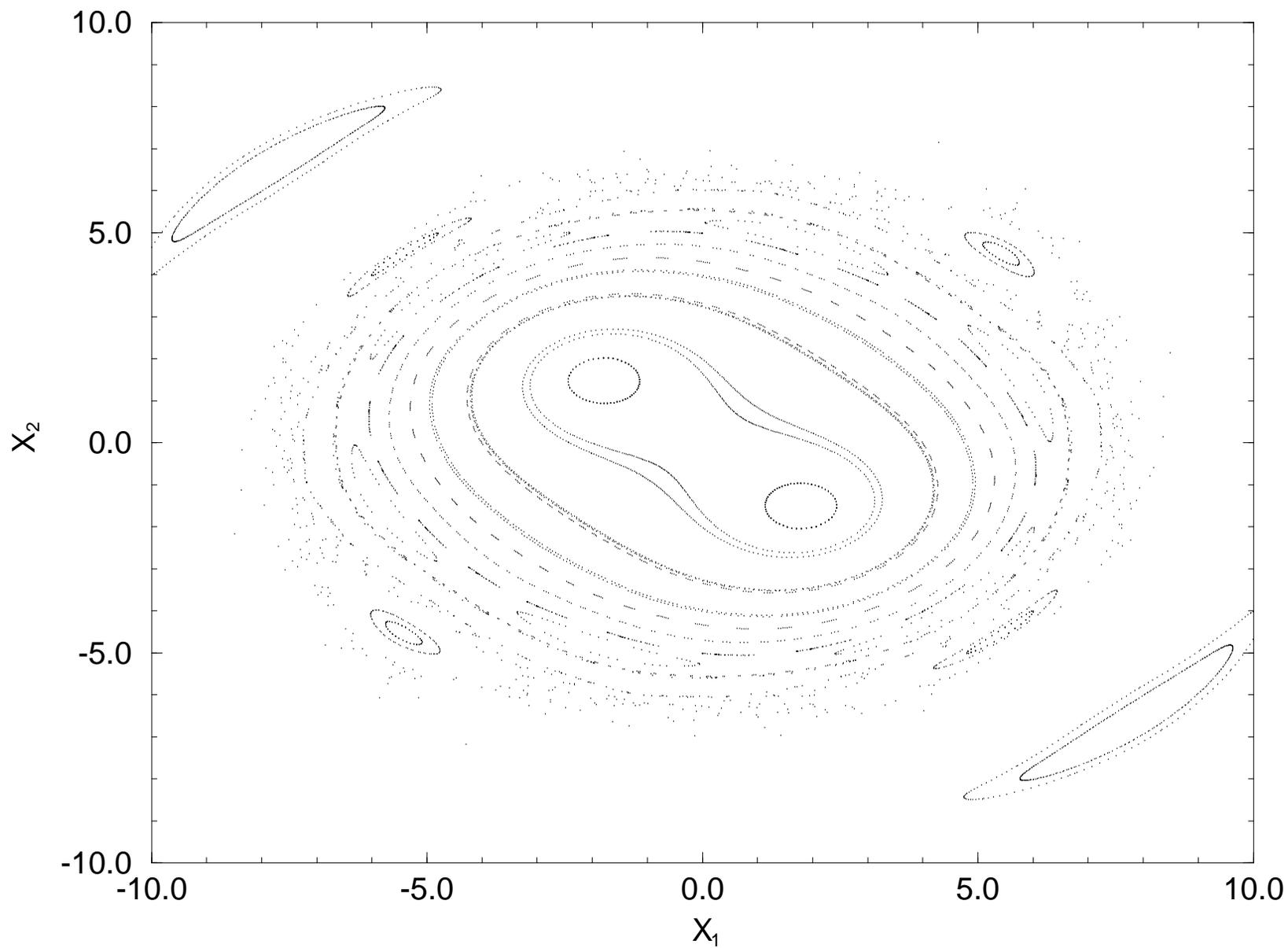
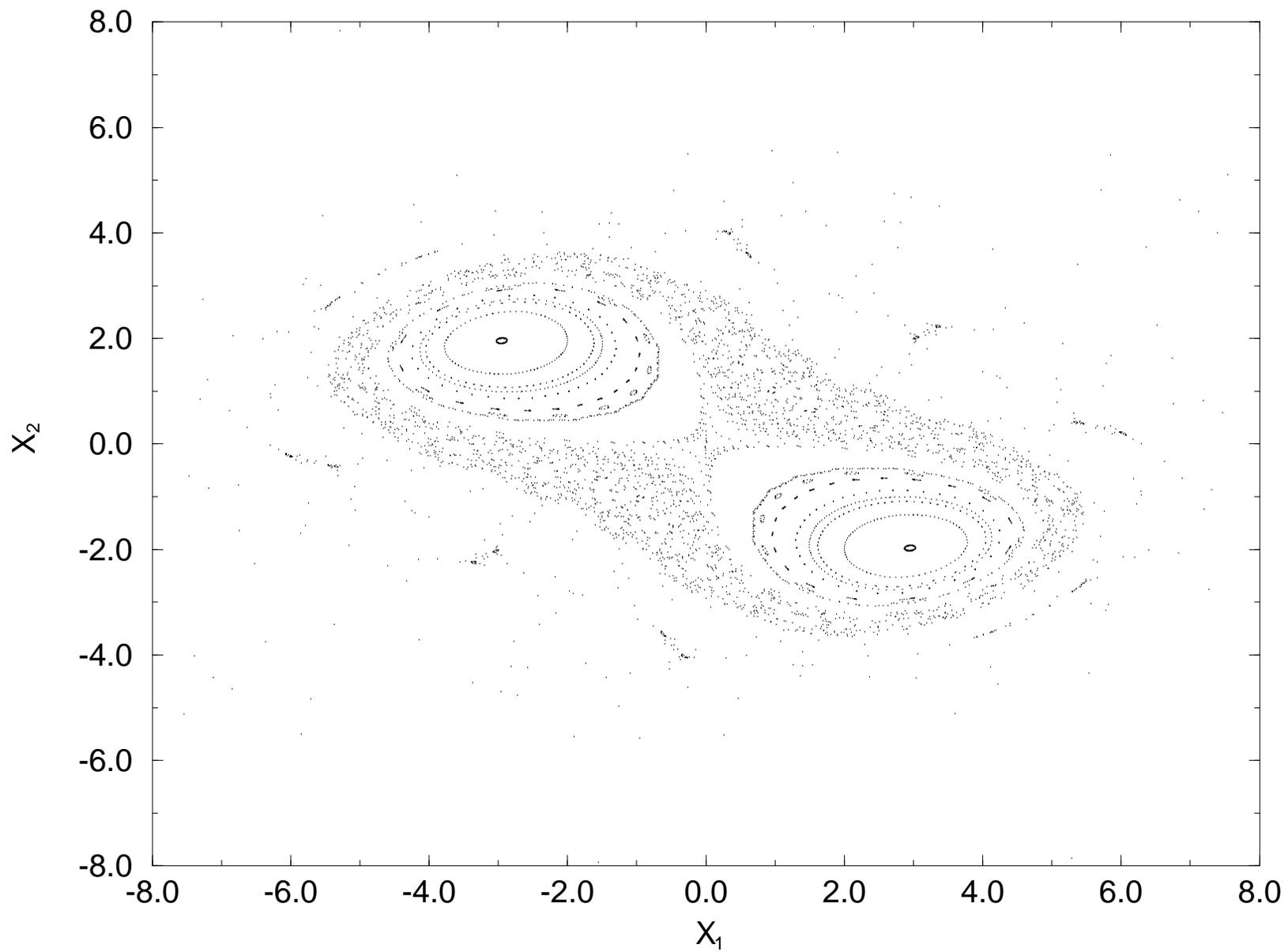
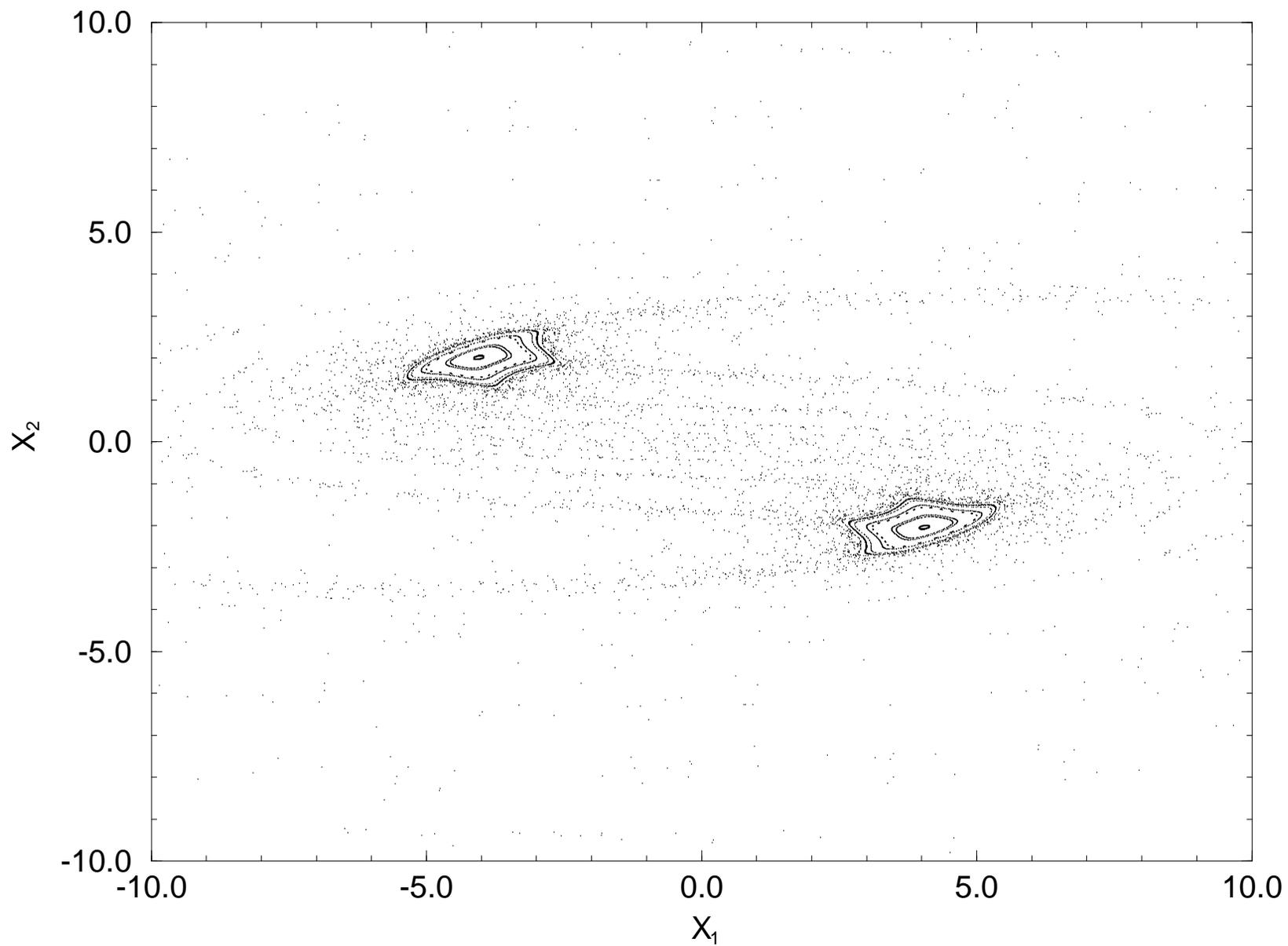
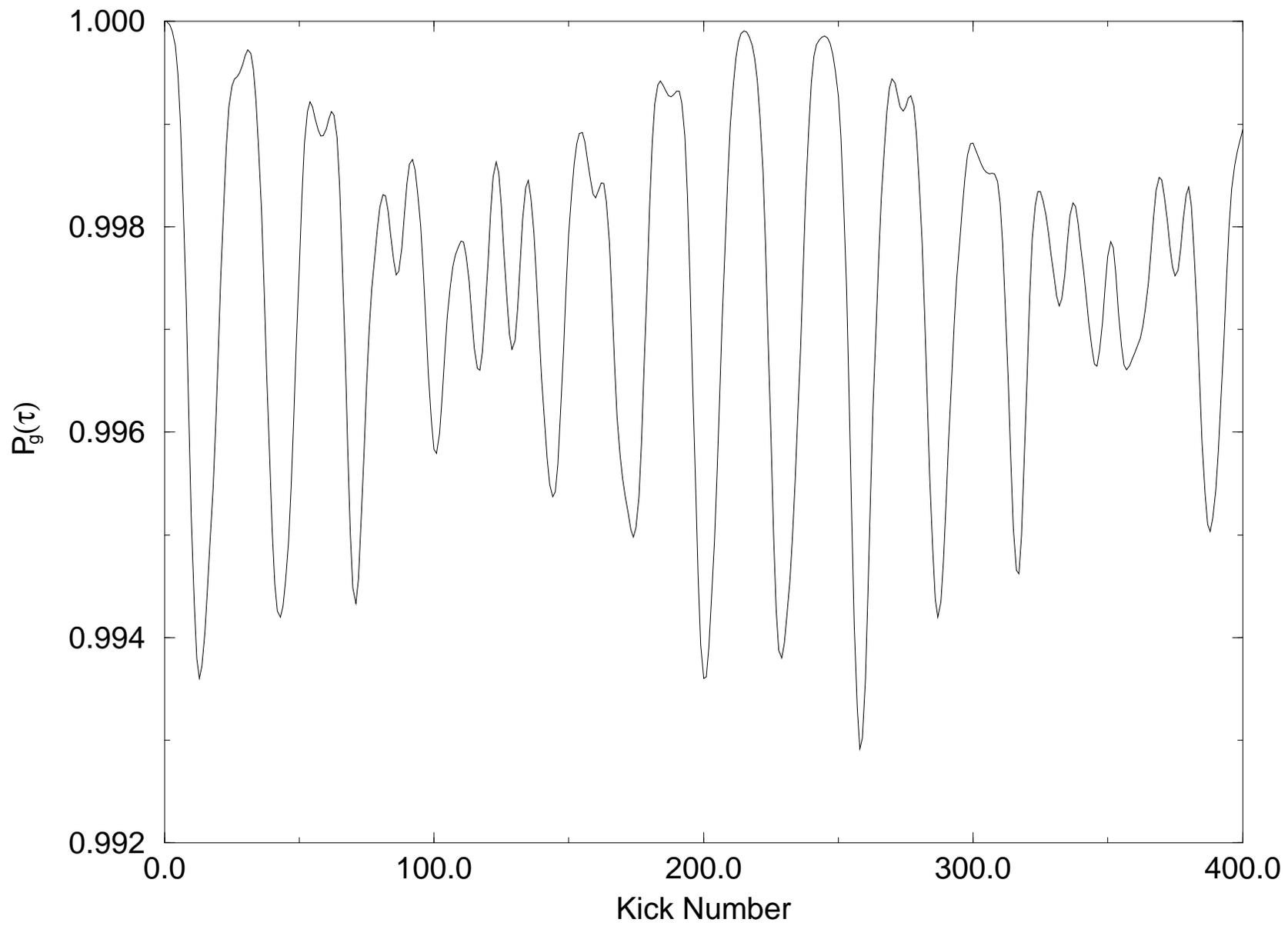
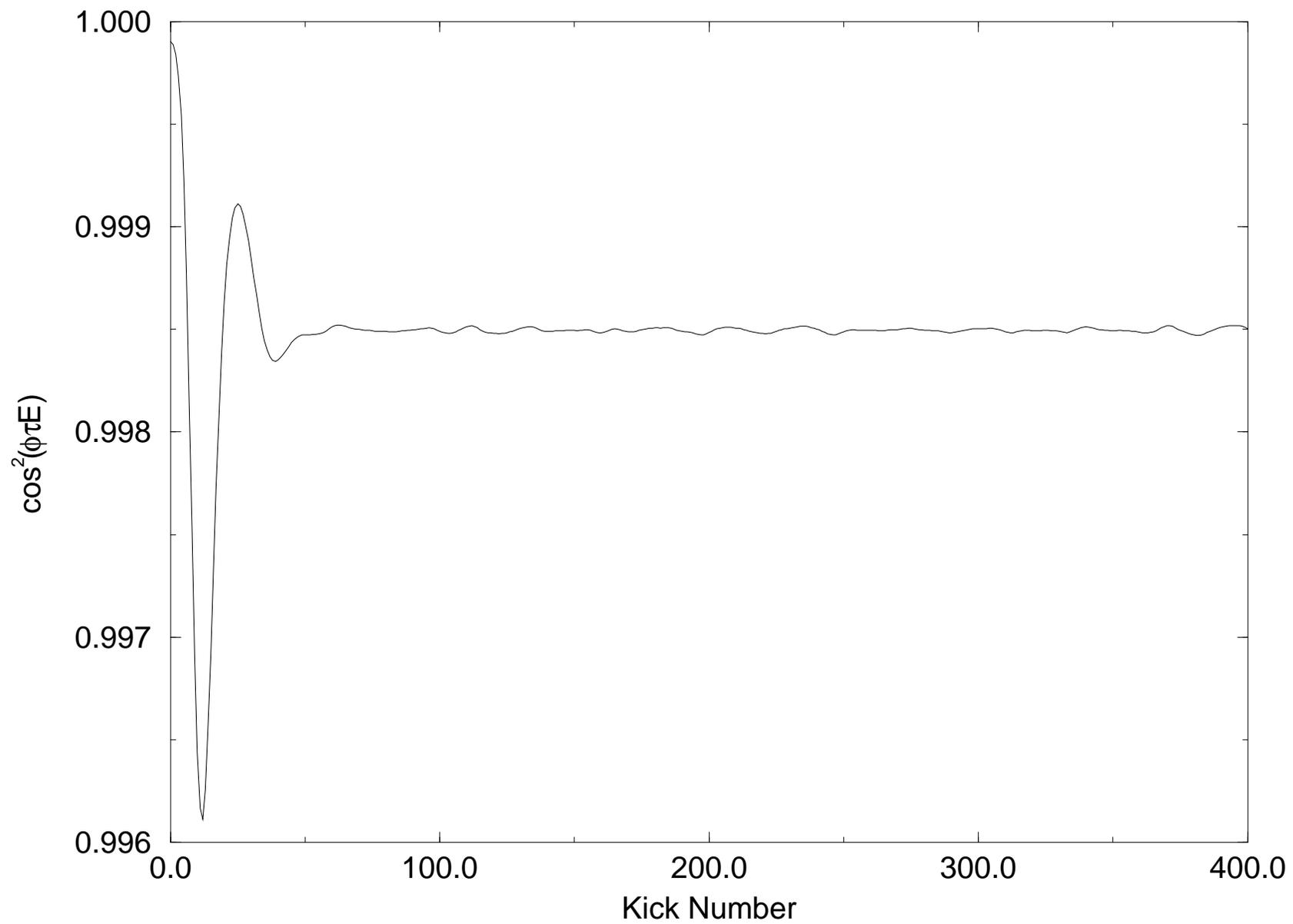
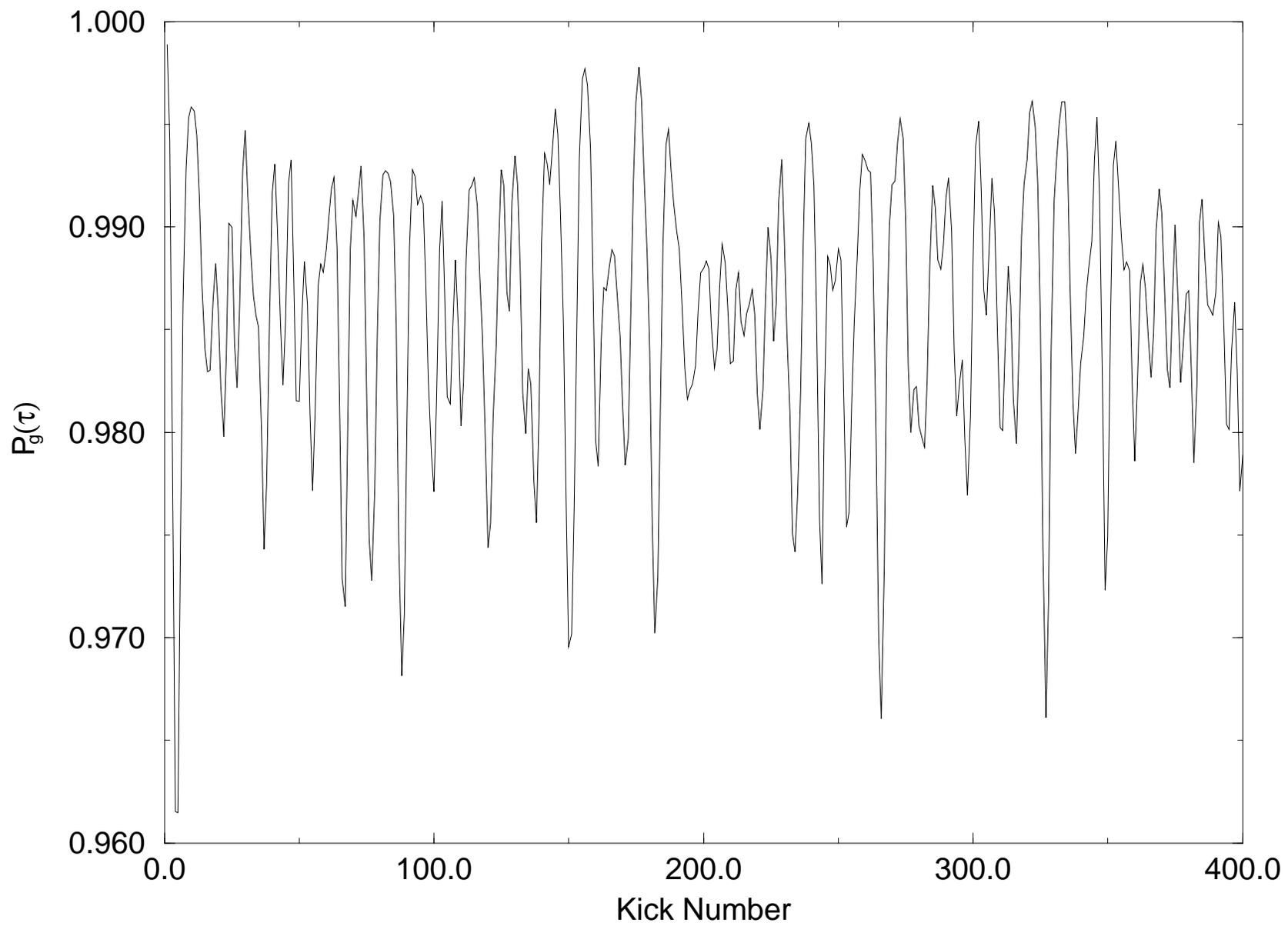
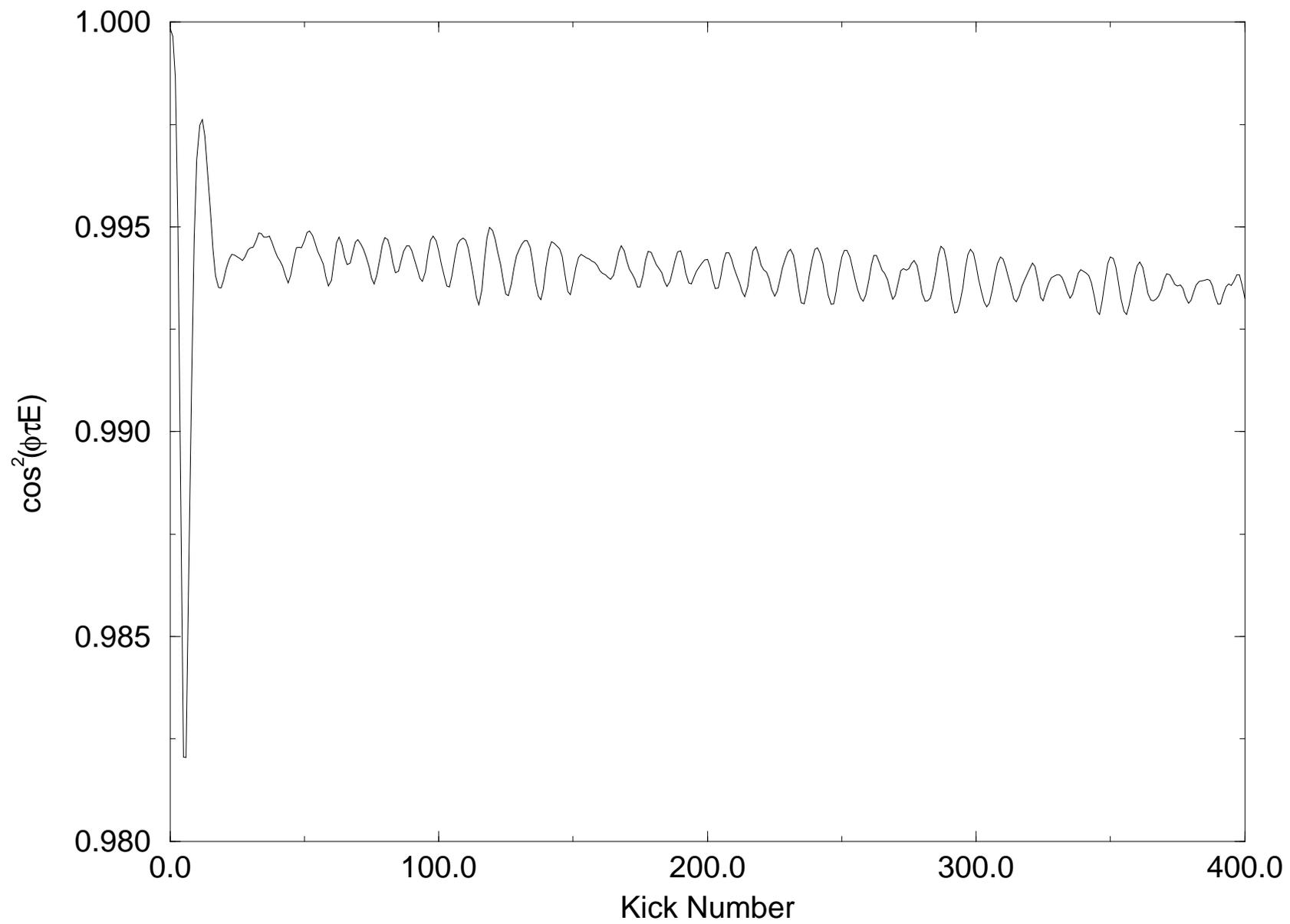
\end{document}